%% file: boundary.tex
\documentclass[12pt]{iopart_mod}

\usepackage{amsfonts,amssymb,amsmath,amsthm,epsfig,cite}

\newcommand{\oB}{\vert_{\partial\mathcal{M}}}

\input{macros}

\DeclareMathOperator{\extdm}{d}
\newcommand{\extd}{\extdm \!}

\newcommand{\macroT}{T}
\newcommand{\macroM}{M}

\eqnobysec 

\begin{document}

\hfill {\footnotesize TUW-05-20}

\hfill {\footnotesize LU-ITP 2005/026}

\title{Physics-to-gauge conversion at black hole horizons}

\author{L.~Bergamin$^1$, D.~Grumiller$^2$, W.~Kummer$^3$, and D.V.~Vassilevich$^4$}

\address{$^{123}$ Institut f\"ur Theoretische Physik, Technische Universit\"at Wien, Wiedner Hauptstr.~8-10/136, A-1040 Vienna, Austria} 

\address{$^{24}$ Institut f\"ur Theoretische Physik, Universit\"at Leipzig, Augustusplatz 10-11, D-04109 Leipzig, Germany} 
\address{$^1$ ESA Advanced Concepts Team (EUI-ACT), ESTEC, Keplerlaan 1,
  NL-2201 AZ Noordwijk, The Netherlands}
\address{$^4$ V.A.~Fock Institute of Physics, St.~Petersburg University, Russia}

\eads{\mailto{bergamin@tph.tuwien.ac.at}, \mailto{grumiller@itp.uni-leipzig.de}, \mailto{wkummer@tph.tuwien.ac.at}, \mailto{Dmitri.Vassilevich@itp.uni-leipzig.de}}

\begin{abstract}
Requiring the presence of a horizon imposes constraints on the
physical phase space. After a careful analysis of dilaton gravity in
2D with boundaries (including the Schwarzschild and Witten black holes
as prominent examples), it is shown that the classical physical phase
space is smaller as compared to the generic case if horizon
constraints are imposed. Conversely, the number of gauge symmetries is
larger for the horizon scenario. In agreement with a recent conjecture
by 't Hooft, we thus find that physical degrees of freedom are converted into gauge degrees of freedom at a horizon.
\end{abstract}


\setcounter{footnote}{0}

\section{Introduction}\label{se:1}

The standard derivation of the Euler-Lagrange equations of motion (EOM) requires boundary conditions on the variation of fields. 
By introducing appropriate boundary terms in the action one can employ different variational principles. The careful treatment of such boundary terms has become a focus of interest in several branches of modern field theory and string theory. In particular, in the context of Black Holes (BHs) edge states have been investigated from various angles (cf.~e.g.~\cite{Balachandran:1991dw}); 
the basic idea of these papers goes back to the seminal one by Witten \cite{Witten:1989hf}. 

The most familiar example where such
issues can be addressed is the Hilbert action of Einstein gravity (EH) in $D$ dimensions in the presence of a (in general non-smooth) boundary, which for Lorentzian signature reads (setting the gravitational coupling $\kappa=1$)
\begin{equation}
  \label{eq:H}
  S^{\rm EH}_D=\frac12 \int\limits_{\mathcal{M}_D}\extd^Dx\sqrt{|g|}R+\!\!\!\!\!\int\limits_{\partial\mathcal{M}_{(D-1)}}\!\!\!\!\!\extd^{(D-1)}x\sqrt{|h|}K+\int\limits_{\partial\partial\mathcal{M}_{(D-2)}}\!\!\!\!\!\!\extd^{(D-2)}x\sqrt{|\sigma|}\alpha\,.
\end{equation}
Here $R$ is the curvature scalar, $K$ the trace of the extrinsic
curvature, $h$ the determinant of the induced metric at the boundary
(the piece-wise smooth part being denoted by $\partial\mathcal{M}$),
$\sigma$ the determinant of the induced metric at each corner and
$\alpha$ the local rapidity (which is the Lorentzian equivalent of a
deficit angle) associated with a given corner. The action \eqref{eq:H}
has been used for instance in \cite{Hayward:1993my,Hawking:1996ww,Booth:1998eh,Park:2001zn}; its boundary terms correspond to the York-Gibbons-Hawking (YGH) boundary term \cite{York:1972sj}, 
the presence of which can be derived e.g.~from the consistency of the variational principle, i.e., by requiring functional differentiability, because the EH action contains second derivatives of the metric.

A fertile field for investigation of gravity always has been its spherical reduction to 2D. In this context
the frequently used second order action \cite{Russo:1992yg} 
of a scalar-tensor theory
\begin{equation}
\label{eq:GDT}
S^{(2)}=\frac12 \int_{\mathcal{M}} \extd^{2}x\, \sqrt{-g}\; \left[ X R + U(X)\; (\nabla X)^{2} - 2V(X)\; \right] \, ,
\end{equation}
encompasses most 2D dilaton gravity models of interest (cf.~e.g.~\cite{Grumiller:2002nm}) including not only spherically reduced Einstein gravity (i.e., the Schwarz\-schild BH) \cite{Berger:1972pg,Kuchar:1994zk,Lau:1996fr} 
but also the Jackiw-Teitelboim model \cite{Barbashov:1979bm}, 
string inspired BH models (including the Witten BH) \cite{Mandal:1991tz} 
and others.
 The functions $U,V$ define the model. The covariant derivative $\nabla$ is associated with the Levi-Civita connection related to the metric $g_{\mu\nu}$, the determinant of which is denoted by $g$. 
In the absence of matter there are no propagating physical degrees of freedom, but the theory nevertheless is not empty: as opposed to \eqref{eq:H} taken at D=2 (which is just the Euler characteristic in the presence of boundaries and deficit angles, cf.~e.g.~\cite{Guillemin:1974}) there are non-trivial EOM and interesting global physical properties, some of which will be recalled in the present work.

Spherical reduction of the action (\ref{eq:H}) leads to a 2D dilaton gravity
action (\ref{eq:GDT}) with the potentials ($\la$ is an irrelevant scale factor and may be set to $1$)
\begin{equation}
V(X)=-\frac{\lambda^2}2 (D-2)(D-3) X^{(D-4)/(D-2)} \,,\quad
U(X)=-\frac{(D-3)}{(D-2)X} \,,\label{eq:SRG}
\end{equation}
supplemented by the boundary term  ($\extd s$ is the arc-length on the boundary)
\begin{equation}
  \label{eq:GDTGH}
  S^{(2)}_{\rm b} = \int_{\partial\mathcal{M}}\extd s X K\,,
\end{equation}
where we assume that corner contributions are absent.
The presence of \eqref{eq:GDTGH} makes the total action functionally
differentiable with respect to the induced metric on the boundary (i.e.~no
normal derivative of the variations of the induced metric appears). This property
does not depend on the potentials $U$ and $V$, and, therefore,
we also adopt (\ref{eq:GDTGH}) for all other dilaton
theories (cf.~e.g.~\cite{Lau:1996fr,Chan:1996sx,Liebl:1997ti}). 
Note that (\ref{eq:GDTGH}) is not just a spherical reduction of the boundary
term in (\ref{eq:H}), as one has to take into account a contribution from
partial integration of the volume term in  (\ref{eq:H}) as well.

In some applications one would like to consider null surfaces as boundary. This is problematic because the extrinsic curvature becomes undefined and thus it is not straightforward to implement e.g.~horizon constraints. In the context of the second order formulation of 4D EH gravity null boundaries were discussed in \cite{Booth:2001gx}, but neither the analysis of the constraint algebra nor the construction of the reduced phase space was performed. As will be shown below this problem may be circumvented in a formalism different from \eqref{eq:H} and \eqref{eq:GDT}. Indeed, there exists a classically equivalent \cite{Katanaev:1995bh} reformulation of \eqref{eq:GDT} as a first order action \cite{Schaller:1994es}, which renders constraint analysis and quantisation surprisingly easy \cite{Grumiller:2002nm}. 
It is one of the prime goals of the present work to translate \eqref{eq:GDTGH} into the first order formulation and to study its impact on the constraint algebra.\footnote{An existing formulation \cite{Kummer:1997si} does not address the main issues of our present work.}
This is not only of interest by itself, but a crucial pre-requisite to address questions regarding quantisation in the presence of a horizon, and one would also expect relevant implications for the BH entropy.

To make this connection more concrete let us briefly review some
recent key results by S.~Carlip \cite{Carlip:2004mn}: 
he
performed a Hamiltonian analysis of 2D dilaton gravity, restricting
the initial/boundary data such that a ``stretched horizon'' is
present.\footnote{Actually, it is not a stretched horizon in the
  technical sense of the word, but some spacelike hypersurface which
  is ``almost null'', i.e., a partial Cauchy hypersurface extending
  from the bifurcation point to lightlike infinity.} Two conditions
are imposed: one sets the Killing norm ``almost'' to zero, the other
one makes the expansion of the hypersurface nearly zero. Although they
are not constraints in the ordinary sense of constrained Hamiltonian
dynamics one may calculate the Poisson brackets between them and it is
found that they convert the
first class constraints generating gauge transformations into second
class constraints. Introducing the Dirac bracket
then establishes the Virasoro algebra with a classical central charge
(which vanishes in the limit of a true horizon). Then, after fixing a
relevant ambiguity, the Cardy formula \cite{Cardy:1986ie} is exploited
to recover the Bekenstein-Hawking relation
\cite{Bekenstein:1973ur}. 
Besides technical issues this
interesting analysis implies an important conceptual question: since
the result seems to be valid for small, but otherwise arbitrary,
``stretching'' of the horizon there appears to be no essential
difference between a horizon as a boundary and some generic spacelike
or timelike boundary. However, if entropy originates from some
microstates attached to the BH horizon one would expect the latter to
play a special role in the analysis. To even address this question it
is therefore necessary to be able to implement ``sharp'' horizon
conditions. It is one of the main tasks of this work to do precisely
that. Horizon boundary conditions will turn out to be quite special
regarding the constraint algebra, the gauge symmetries and the
physical phase space. Our result turns out to be concurrent with a
recent idea by 't Hooft \cite{'tHooft:2004ek}. 

This paper is organised as follows: in section \ref{se:2} the first
order formulation of 2D dilaton gravity is recapitulated
briefly. Boundary terms are introduced in section \ref{se:3}. Then the
constraint algebra (section
\ref{se:4}) as well as the gauge symmetries (section \ref{se:5}) are
analysed in detail. The classical physical phase space is studied in section
\ref{se:6},  while the final section \ref{se:7} concludes with an extensive
discussion of our results and an outlook to generalisations and
applications. 

\section{First order formulation}\label{se:2}

Already classically, but especially at the quantum level,
it is very convenient in 2D to employ the first order formalism in terms of the ``Cartan variables'' Zweibeine $e^a$ and spin connection $\om$.\footnote{In our notation $e^a=e^a_\mu \extd x^\mu$ is the
dyad 1-form. Latin indices refer to an anholonomic frame, Greek indices to a holonomic one.
The 1-form
$\omega$ represents the  spin-connection $\om^a{}_b=\epsilon^a{}_b\om$
with  the totally antisymmetric Levi-Civita symbol $\epsilon_{ab}$ ($\epsilon_{01}=+1$). With the
flat metric $\eta_{ab}$ in light-cone coordinates
($\eta_{+-}=1=\eta_{-+}$, $\eta_{++}=0=\eta_{--}$) it reads $\epsilon^\pm{}_\pm=\pm 1$. The Levi-Civita symbol with anholonomic indices is fixed as $\tilde{\epsilon}_{01}=-\tilde{\epsilon}^{01}=1$. The metric is determined by $g_{\mu\nu}=e^+_\mu e^-_\nu+e^-_\mu e^+_\nu$. The determinant of the Zweibein is denoted by $e=\det e^a_\mu=e_0^-e_1^+-e_0^+e_1^-$. 
The torsion 2-form is given by 
$T^\pm=(\extd\pm\omega)\wedge e^\pm$. 
The curvature 2-form $R^a{}_b$ can be represented by the 2-form $\mathcal{R}$ defined by 
$R^a{}_b=\epsilon^a{}_b \mathcal{R}$, $\mathcal{R}=\extd\om$. The volume 2-form is denoted by $\epsilon = e^+\wedge e^-$. 
The overall minus sign in \eqref{eq:FOG} is the only difference to the notation used in \cite{Grumiller:2002nm}. 
}
The first order gravity action \cite{Schaller:1994es}
\eq{
S^{\rm (1)}=-\int_{\mathcal{M}}  \left[X_aT^a+X\mathcal{R}+\epsilon\mathcal{V} (X^aX_a,X)\right]
}{eq:FOG}
encompasses essentially all known dilaton theories. 
The fields $X$ and $X_a$ are Lagrange multipliers for curvature $\mathcal{R}$ and torsion $T^a$, respectively. The former coincides with the dilaton field in the second order formulation.
Actually, for most practical purposes the special case
\begin{equation}
  \label{eq:pot}
  \mathcal{V} (X^aX_a,X) = X^+X^- U(X) + V(X)
\end{equation}
is sufficient because \eqref{eq:FOG} with \eqref{eq:pot} is equivalent to the second order action \eqref{eq:GDT} with the same potentials $U,V$. 
For $U\neq 0$ the action \eqref{eq:FOG} implies non-vanishing torsion. Therefore also $\mathcal{R}$ should not be confused with the Hodge dual of $R$ in \eqref{eq:GDT} where the connection is torsionless (and metric compatible).
It is useful to introduce the notation
\begin{equation}
  \label{eq:Q}
  Q(X)=\int^X U(y)\extd y
\end{equation}
and
\begin{equation}
  \label{eq:w}
  w(X)=\int^X V(y) \exp{Q(y)} \extd y\,.
\end{equation}
The latter combination of the potentials $U,V$ remains invariant under local Weyl rescalings $g_{\mu\nu}\to\Om^2(X)g_{\mu\nu}$. Both definitions contain an ambiguity from the integration constant which may be fixed conveniently.

It has been pointed out in \cite{Schaller:1994es} that \eqref{eq:FOG} is a particular Poisson-$\si$ model (PSM), 
\begin{equation}
  \label{eq:PSM}
  S^{\rm PSM}= - \int_{\mathcal{M}} \left[ X^I \extd A_I - \frac12 P^{IJ} A_J\wedge A_I \right]
\end{equation}
with a three-dimensional target space, the coordinates of which are $X^I=\{X,X^+,X^-\}$. The gauge fields comprise the Cartan variables, $A_I=\{\om,e^-,e^+\}$.
Because the dimension of the Poisson manifold is odd the Poisson tensor 
\begin{equation}
  \label{eq:Ptensor}
  P^{X\pm} =\pm X^\pm\,, \qquad
  P^{+-} = {\mathcal V}\,, \qquad
  P^{IJ} = -P^{JI}\,,
\end{equation}
cannot have full rank. Therefore, always a Casimir function,
\begin{equation}
  \label{eq:C}
  \mathcal{C} = X^+X^-\exp{Q(X)} + w(X)\,,
\end{equation}
exists whose absolute conservation,
\begin{equation}
  \label{eq:conslaw}
  \extd \mathcal{C}=0\,,
\end{equation} 
has been investigated extensively
\cite{Frolov:1992xx,Grosse:1992vc}. 
For spherically reduced gravity $\mathcal{C}$ is related directly to the ADM
mass.

In the context of 2D dilaton gravity equation \eqref{eq:conslaw} is
sometimes called ``generalised Birkhoff theorem'' because from the exact
solution for the line element
(cf.~e.g.~\cite{Klosch:1996fi,Grumiller:2002nm}) following from
\eqref{eq:FOG} with \eqref{eq:pot} it is evident that there is always
a Killing vector $k=k^\mu\partial_\mu$ with the norm
\begin{equation}
  \label{eq:killingnorm}
  k^\mu k_\mu=2X^+X^-\exp{Q(X)}\,.
\end{equation}
This implies that the condition for a Killing horizon may not only be imposed on the worldsheet,\footnote{In \eqref{eq:killinghorizonws} it has been assumed that the Killing horizon is a surface of constant coordinate $x^1$ and therefore the Killing norm is proportional to $e_0^-e_0^+$. This can always be achieved by a suitable choice of coordinates.}
\begin{equation}
  \label{eq:killinghorizonws}
  {\rm Killing\,\,horizon\,\,(world\,\,sheet):}\quad e_0^-e_0^+=0\,,
\end{equation}
but alternatively also in target space,
\begin{equation}
  \label{eq:killinghorizon}
  {\rm Killing\,\,horizon\,\,(target\,\,space):}\quad X^+X^-=0\,.
\end{equation}
Note that on-shell \eqref{eq:C} allows to express the Killing norm as a function of $X$ and the value of the Casimir function. 
In axial gauge ($e_1^-=1, e_1^+=0$) the metric simplifies to
\begin{equation}
  \label{EFmetric}
  g_{\mu\nu}^{\rm (EF)}=e_0^+\left(\begin{array}{cc}
 2e_0^- & 1\\
 1 & 0
\end{array}\right)_{\mu\nu}
\end{equation}
and the (world-sheet) condition for a Killing horizon reads $e_0^-=0$.
 The first EOM in equation \eqref{eq:EOM} then implies $X^-=0$ since
 the dilaton is constant on the horizon. If additionally the dilaton
 is assumed to depend on $x^1$ only (which is always possible in the
 absence of matter and within a certain patch) then the gauge implied by \eqref{EFmetric} is of Eddington-Finkelstein type. We do not go further into details explaining 2D dilaton gravity without boundaries but refer to the extensive review \cite{Grumiller:2002nm}, upon which this work is based.

Subsequently we will often drop the attribute ``Killing'' as it is
understood that all horizons studied in the present work are Killing
ones on-shell. However, it is emphasised that our results below should
generalise to trapping horizons \cite{Hayward:1993wb}, which are defined by the condition \eqref{eq:killinghorizon} but not \eqref{eq:killinghorizonws}. This point will be addressed in more detail in the conclusions where systems with matter are outlined.

\section{Boundary terms in 2D dilaton gravity}\label{se:3}

As pointed out in the previous section the first order formulation
of 2D dilaton gravity has many advantages over the second order 
formulation. In this section we rewrite the boundary term (\ref{eq:GDTGH})
in a first order form and discuss the boundary conditions which 
follow from varying the total action.
First we have to find an alternative form of the extrinsic curvature
which involves the connection $\omega$.

To clarify the relation between $\omega$, the Levi-Civita connection $\hat{\om}$ and the extrinsic curvature,
consider the Euclidean case for simplicity. Recalling the main definitions
(see e.g.\ \cite{Gilkey:1994}), let $\mathcal{M}$
be a manifold of dimension $D$, let $\bar e_\bot$ be an inward pointing unit
vector on the boundary, let $\bar e_i$ ($i=1,\dots,D-1$) be a local
orthonormal frame for the tangent bundle of $\partial\mathcal{M}$.
If one identifies $\bar e$ with our dynamical Vielbein $e$, this would
mean partially fixing the gauge. Then the extrinsic curvature is given by $K_{ij}=\bar{\omega}^\bot{}_{ij}$ and its trace is denoted by $K$.
Let us denote by a semicolon ($;$) multiple covariant derivatives with respect to the
Levi-Civita connection $\bar \omega$ (associated with $\bar e$), 
and by a colon ($:$) multiple covariant
derivatives with respect to the Levi-Civita connection on the boundary.
The extrinsic curvature tensor measures the difference between these two
connections:
\begin{equation}
v_{i;j}=v_{i:j}-K_{ij}v_\bot \,,\label{vij}
\end{equation}
where $v$ is a co-vector field on $\mathcal{M}$. We denote $i,j,\dots$ by $\|$.
For a two-dimensional manifold $\partial{\mathcal{M}}$ is one-dimensional.
 Since $so(1)$ is trivial, 
\begin{equation}
v_{\|:\|}=\partial_\| v_\|\,.
\end{equation}
On the other hand, the expression
\begin{equation}
v_{\|;\|}=\partial_\| v_\| -\epsilon^\bot{}_\|\bar{\omega}_\| v_\bot
\end{equation}
by virtue of (\ref{vij}) yields
\begin{equation}
K_{\|\|}=\epsilon^\bot{}_\|\bar{\omega}_\| \label{ktt}\,.
\end{equation}
It is recalled that $\epsilon_{\|\bot}=\tilde{\epsilon}_{01}=1$ if the boundary is assumed to be at $x^1=\rm const$.

The connections $\bar\omega$ and $\hat\omega$ are related by a Lorentz
transformation. Performing it after
returning to Minkowskian signature 
obtains
\begin{equation}
K=\hat \omega_\| +\frac 12
\partial_\|\ln{\left|\frac{e_\|^+}{e_\|^-}\right|}\, ,
\label{Kom}
\end{equation}
where $e_\|^\pm$ means again the component of the
Zweibein parallel to the boundary.
In terms of the full spin-connection $\om$ the first order version of the boundary term \eqref{eq:GDTGH} reads
\begin{equation}
  \label{eq:bound2}
\int_{\partial\mathcal{M}} \left(X\omega
+\frac 12  X\extd\,\ln{\left|\frac{e_\|^+}{e_\|^-}\right|}\right)\,.
\end{equation}
It has been obtained also in \cite{Kummer:1997si}, although from quite a different argument.
A local Lorentz transformation (with infinitesimal or finite Lorentz angle $\ga$),
\begin{equation}
  \label{eq:loclor}
  \om^\prime=\om-\extd\ga\,,\quad (e^\pm)^\prime=e^\pm\exp{[\pm\ga]}\,,
\end{equation}
indeed leaves \eqref{eq:bound2} invariant. 
For sake of definiteness from now on it will be supposed that $\partial\mathcal{M}$ is the \emph{lower} boundary in $x^1$ direction and that there is no boundary in $x^0$ direction. Then the full action 
reads
\begin{align}
S&=-\int_{\mathcal{M}} \extd^2x \left[ X_a (D_\mu e_\nu^a) \tilde\epsilon^{\mu\nu}
+X\partial_\mu \omega_\nu \tilde \epsilon^{\mu\nu}
+ e(UX^+X^- +V) \right] \nonumber \\
&\quad -\int_{\partial{\mathcal{M}}} \extd x^0 
\left[ X \omega_0 +\frac 12 X \partial_0 \ln \left(
\frac {e_0^+}{e_0^-} \right) \right]\equiv \int \extd x^0 \mathcal{L}
 \,.\label{act1}
\end{align}
We have dropped the absolute value in the last boundary term because we will exclusively discuss patches where the sign of the ratio $e_0^+/e_0^-$ is (semi-)positive.

To prove that this action is indeed equivalent to the second order action
(\ref{eq:GDT}) with (\ref{eq:GDTGH}) one has to use twice the algebraic
EOM for $X^a$ \cite{Katanaev:1997ni,Grumiller:2002nm}.
It is crucial that the variation of (\ref{act1}) with respect to $X^a$
does not produce any boundary terms. Therefore, the proof of 
 \cite{Katanaev:1997ni,Grumiller:2002nm} does not require any essential
modifications due to the presence of boundaries.

The variation of (\ref{act1}) produces the EOM in the bulk \eqref{eq:EOM}
and the boundary terms
\begin{multline}
\int_{\partial{\mathcal{M}}} \extd x^0 
\left[ (\delta e_0^-) \left( X^+ -\frac{\partial_0 X}{2e_0^-} \right)
+(\delta e_0^+) \left( X^- +\frac{\partial_0 X}{2e_0^+} \right)\right. \\
- \left. (\delta X) \left( \omega_0 + \frac 12 \partial_0
\ln \left(
\frac {e_0^+}{e_0^-} \right)\right) \right] \,.\label{bvari}
\end{multline}
We shall exclusively consider boundaries on which the value of the dilaton
is fixed to a constant,
\begin{equation}
X\oB = {\rm const.}\quad\Rightarrow\quad \partial_0X\oB=\de X\oB=0\,.\label{DbconX}
\end{equation}
Then to cancel (\ref{bvari}) it is sufficient to impose 
\begin{equation}
  \label{eq:bc3}
 X^+\de e^-_0 \oB=0\,,\qquad
 X^-\de e^+_0 \oB=0\,. 
\end{equation}
All boundary conditions which we consider in this paper do satisfy
(\ref{DbconX}) and (\ref{eq:bc3}). Three cases are of particular interest.
Fixing $e_0^\pm\oB$ corresponds to a \emph{generic boundary}. The requirement
$X^\pm\oB=0$ will be called \emph{bifurcation point boundary condition} because on-shell it holds at the bifurcation point on the Killing horizon. And, finally, $X^-\oB =0$, $e_0^-\oB =0$ defines
a  \emph{horizon boundary condition} (cf.~\eqref{eq:killinghorizon}).
We stress that the condition $e_0^-\oB =0$ is possible only 
if we also fix the dilaton to a constant on $\partial \mathcal{M}$,
because otherwise the action \eqref{act1} would become singular.

So far we have disregarded the possibility of corners. In their presence \eqref{act1} has to be replaced by a similar action, but with the last boundary term partially integrated (cf.~section V.~of \cite{Kummer:1997si}, (5.1)-(5.6)),
\begin{equation}
  S^{\rm tot}=S^{\rm (1)} + \int\limits_{\partial\mathcal{M}_s}\extd s XK + \sum X\alpha\,,
\label{eq:nolabel}
\end{equation}
where $K$ is the extrinsic curvature with respect to the full spin-connection. Here $\partial\mathcal{M}_s$ extends over the smooth parts of the full
boundary and the sum is over the corner points; $\alpha$ 
is again the local rapidity at each corner. The
boundary terms in \eqref{eq:nolabel} contain a multiplicative factor
$X$ in comparison with the ones in \eqref{eq:H}. 
If $\de X=0$ at the corners then no additional terms are produced by variation of \eqref{eq:nolabel} as compared to \eqref{bvari}. 
Subsequently smoothness of $\partial\mathcal{M}$ will be assumed 
and hence corner terms will play no role.

Within the first order formulation there are other choices of boundary
terms which may look natural. For example, instead of the second line in
(\ref{act1}) one can use
\begin{equation}
  \label{eq:bc2}
-\int_{\partial{\mathcal{M}}} \extd x^0 
\left[ X \omega_0 + X^+e_0^- + X^-e_0^+ \right] .
\end{equation}
This prescription has been studied by Gegenberg, Kunstatter 
and Strobl \cite{Gegenberg:1997de} (cf.~also \cite{Strobl:1999wv}). 
However, in this approach the action is not invariant under the (unrestricted)
Lorentz transformations, and, therefore, the corresponding first order
model cannot be equivalent to a second order action which is formulated
in terms of Lorentz invariant fields (the metric and the dilaton). 
On a more technical side, the variation of the action \eqref{eq:bc2}
with respect to $X^\pm$ produces some boundary terms which violate the
standard proof of the equivalence between the two formulations of dilaton
gravity \cite{Katanaev:1997ni}. As we would like to make contact with
the second order formulation and the standard YGH result we shall
concentrate on the action \eqref{act1}.

\section{Constraint algebra in presence of boundaries}\label{se:4}

In the Hamiltonian approach the boundary conditions become
constraints. Therefore, one has to analyse the constraint 
algebra which includes these new constraints, separate second
class ones, and define the Dirac bracket. Some
general aspects of this procedure were developed in 
\cite{Solovev:1993zf} 
where one can also find further references.
More recently this approach was applied to Dirichlet branes
\cite{Ardalan:1999av}. 

Typically, in the presence of boundaries all gauge symmetries
are broken, i.e., the action and the boundary conditions
for the fields are gauge invariant only if the parameters of gauge transformations are
restricted at $\partial\mathcal{M}$.
Sometimes even such partial gauge invariance is encountering serious obstacles
(cf. \cite{vanNieuwenhuizen:2005kg} where supersymmetric boundary
conditions for supergravity were analysed). In this respect the
situation with the gauge symmetries we observe below is quite
specific. If the boundary corresponds to the BH horizon,
one does not need to impose conditions on some of the gauge
parameters to achieve full gauge invariance of the action and of
the boundary conditions. This may be interpreted as a manifestation of 
gauge degrees of freedom localised on the horizon whose
existence  was proposed
recently by 't Hooft \cite{'tHooft:2004ek}.

The full set 
of the boundary conditions for the fields and for the gauge parameters
requires several consistency conditions. Boundary terms in the
Euler-Lagrange and in the symmetry variation of the action should vanish, and the set of the boundary conditions
on the fields should be closed under gauge transformation.
All these requirements are satisfied in our scheme, so that
we actually construct an {\em orbit of boundary conditions} 
in the sense of \cite{Lindstrom:2002mc}.\footnote{This notion was
originally introduced to study supersymmetry in two dimensions, and 
its practical use included quantum corrections to the mass of
supersymmetric solitons \cite{Bordag:2002dg}. 
Gauge invariant boundary conditions for four-dimensional perturbative
quantum gravity were constructed in 
\cite{Barvinsky:1987dg}. 
In Euclidean 2D $R^2+T^2$ gravity the corresponding analysis of boundary 
conditions and one-loop divergences
was performed in \cite{Vassilevich:1995zk}. In that paper it was demonstrated
that the volume divergences are cancelled (in accordance with the absence
of bulk degrees of freedom), but the boundary divergences are not. This
result indicates that some boundary degrees of freedom
are present in the model.}

\subsection{No boundaries: a brief review}

Because appropriate generalisations are needed in the presence of
boundaries, and also to fix our notation, we shortly review the
constraint analysis in the absence of boundaries in the first order
formulation \eqref{eq:FOG}. By $\{,\}$ we always mean the Poisson
bracket. Expressions of the type $\{q,p^\prime\}$ imply that $q$ is
taken at point $x^1$ and $p$ at point ${x^1}^\prime$. The short hand
notation $\de$ for $\de(x^1-{x^1}^\prime)$ is used. No distinction
need be made between upper and lower canonical indices, i.e., we will employ exclusively
\begin{equation}
  \label{eq:canbrac}
  \{q_i,p_j^\prime\}=\de_{ij}\de\,,\qquad\{q_i,q_j^\prime\}=0=\{p_i,p_j^\prime\}\,.
\end{equation}
It is supposed that $x^0$ is our ``time'' variable in the Hamiltonian formulation, although it is emphasised that $x^0$ might as well be a radial or lightlike coordinate. Whatever the physical interpretation of the coordinate $x^0$, it is the quantity with respect to which the Hamiltonian generates translations.
 From the first line in the action \eqref{act1} one can see immediately that the target space coordinates $X,X^\pm$ are canonically conjugate to the $1$-components of the gauge fields,
\begin{align}
\label{eq:coo}
& {\rm coordinates:}\,\, && (q_1,q_2,q_3)=(\om_1,e_1^-,e_1^+)\,, && (\bar{q}_1,\bar{q}_2,\bar{q}_3)=(\om_0,e_0^-,e_0^+)\,,\\
& {\rm momenta:}\,\, && (p_1,p_2,p_3)=(X,X^+,X^-)\,, && (\bar{p}_1,\bar{p}_2,\bar{p}_3)\,,
\label{eq:mom}
\end{align}
whereas the $0$-components of the gauge fields encounter no canonically conjugate partner. Consequently, primary first class constraints 
\begin{equation}
  \label{eq:primaryfirstclassconstraints}
  \bar{P}_i=\bar{p}_i\approx0
\end{equation}
are produced ($\approx 0$ means ``weakly vanishing'').
They generate secondary first class constraints $G_i\approx 0$, where
\begin{align}
  \label{eq:G1}
 & G_1=\partial_1p_1+p_3q_3-p_2q_2\,,\\
  \label{eq:G2}
 & G_2=\partial_1p_2+q_1p_2-q_3\mathcal{V}(p_2p_3,p_1)\,,\\
  \label{eq:G3}
 & G_3=\partial_1p_3-q_1p_3+q_2\mathcal{V}(p_2p_3,p_1)\,.
\end{align}
As expected for a reparametrisation invariant theory, the Hamiltonian density is a sum over constraints, $\mathcal{H}=-\bar{q}_iG_i$. While all Poisson brackets between any of the $\bar{P}_i$ with any other constraint vanish trivially, the brackets between the secondary first class constraints yield the algebra
\begin{equation}
  \label{eq:consalg}
\{G_i,G_j^\prime\}=C_{ij}{}^kG_k\de\,.
\end{equation} 
It is one of the remarkable features of the first order formulation
\eqref{eq:FOG} that the constraint algebra closes with delta functions and not with derivatives
thereof. Thus it resembles a Lie-algebra, albeit it is non-linear, i.e., there are structure functions $C_{ij}{}^k(p_l)=-C_{ji}{}^k(p_l)$ whose non-vanishing components read
\begin{equation}
  \label{eq:strucfunc}
C_{12}{}^2=-1\,,\quad C_{13}{}^3=+1\,,\quad C_{23}{}^i = -\frac{\partial\mathcal{V}}{\partial p_i}  \,,
\end{equation}
and whose centre is generated by the Casimir function $\mathcal{C}$ as
defined in
\eqref{eq:C} and $\partial_1\mathcal{C}$ \cite{Grosse:1992vc}. Since
the structure functions only depend on $p_i$ and because the Poisson
brackets of $p_i$ with $G_j$ also yield some functions of $p_i$ alone one obtains a finite W-algebra if the $p_i$ are included in the set of generators of the algebra.
For details on such algebras cf.~e.g.~\cite{deBoer:1995nu}.
Physically, the constraint $G_1$ may be identified as the generator of
local Lorentz transformations, while (certain combinations of) $G_2$
and $G_3$ constitute the two diffeomorphism constraints. In the
presence of minimally coupled matter\footnote{``Minimally coupled''
  means no coupling to the dilaton field -- for instance, a minimally
  coupled massless scalar field $\phi$ has the  Lagrange density
  $\sqrt{-g}(\nabla\phi)^2$.} the constraints are modified, but their algebra is not. For non-minimally coupled matter the structure function $C_{23}{}^1$ acquires an additional, matter dependent, contribution \cite{Grumiller:2001ea}.

Because classically \eqref{eq:FOG} is equivalent to \eqref{eq:GDT} it is possible to recover the Virasoro algebra by appropriate recombinations of the constraints $G_i$. Indeed, by taking the linear combinations $\mathcal{E}=q_1G_1-q_2G_2+q_3G_3$, $\mathcal{P}=q_iG_i$ the new constraints $\mathcal{E}$, $\mathcal{P}$ fulfil \cite{Katanaev:1993fu}
\begin{align}
 \{\mathcal{E},\mathcal{E}^\prime\} &=(\mathcal{P}+\mathcal{P}^\prime)\partial_1\de\,, \label{eq:vir1}\\
 \{\mathcal{P},\mathcal{P}^\prime\} &=(\mathcal{P}+\mathcal{P}^\prime)\partial_1\de\,, \label{eq:vir2}\\
 \{\mathcal{E},\mathcal{P}^\prime\} &=(\mathcal{E}+\mathcal{E}^\prime)\partial_1\de\,. \label{eq:vir3}
\end{align}
In the presence of a central charge $c$ the last Poisson bracket acquires an additional term $ic/(12\pi)\partial_1^3\de$ on the right hand side. We conclude this brief review by remarking that another linear combination of the $G_i$ is possible which makes the constraint algebra abelian (not only locally, which is always possible \cite{Henneaux:1992} of course, but in a certain patch). This is discussed in detail in \ref{app:A}.

\subsection{Modifications due to boundaries}

It is recalled that exclusively the prescription \eqref{DbconX}, \eqref{eq:bc3} is employed. We start the analysis by rewriting \eqref{act1} in terms of canonical variables \eqref{eq:coo}-\eqref{eq:mom},\footnote{As before $\partial{\mathcal{M}}$ refers to the {\em lower} boundary in $x^1$ direction. Note that this implies some unusual minus signs in partial integration, e.g., $\int y\partial x=-\int x\partial y -xy|_{\partial{\mathcal{M}}}$.}
\begin{equation}
S=\int_{\mathcal{M}} \extd^2x \left[ p_i \partial_0 q_i + \bar q_i G_i^{\rm bulk} \right]
+\int_{\partial{\mathcal{M}}} \extd x^0\left[
p_2 \bar q_2 + p_3 \bar q_3 -\frac 12 p_1 \partial_0 
\ln \left( \frac {\bar q_3}{\bar q_2} \right) \right] \,,\label{act2}
\end{equation}
with $G_i^{\rm bulk}$ equal to $G_i$ as defined in \eqref{eq:G1}-\eqref{eq:G3}.
It is useful to smear the constraints with test functions $\eta(x^1),\xi(x^1),\dots$ For the primary ones this yields (cf.~\eqref{eq:primaryfirstclassconstraints})
\begin{align}
& \bar{P}_1[\eta] = \int \extd x^1\bar{p}_1\eta\,,\label{eq:p1}\\
& \bar{P}_2[\eta] = \int \extd x^1\bar{p}_2\eta 
-\frac12 \frac{p_1}{\bar{q}_2}\eta\oB \,,\label{eq:p2}\\
& \bar{P}_3[\eta] = \int \extd x^1\bar{p}_3\eta
+\frac12 \frac{p_1}{\bar{q}_3}\eta\oB \label{eq:p3}\,.
\end{align}
The total Hamiltonian reads
\begin{equation}
H^{\rm tot} =  \int \extd x^1 (p_i \partial_0 q_i +\bar p_i \partial_0 \bar q_i )-
\mathcal{L} + \sum_i \bar{P}_i[\lambda_i] \,,
\label{totHam}
\end{equation}
and the canonical one is defined as
\begin{equation}
H = -\int \extd x^1\, \bar q_i G_i^{\rm bulk} -
 (
p_2 \bar q_2 + p_3 \bar q_3 )\oB \label{Ham1} \,.
\end{equation}
As usual $H^{\rm tot} = H + \sum_i\bar{P}_i[\tilde{\lambda}_i]$ with
$\tilde{\lambda}_i = \lambda_i+\partial_0 \bar{q}_i$.
The Poisson brackets between $H$ and the primary constraints generate 
secondary ones,
\begin{equation}
\{\bar P_i[\eta] ,H\} = G_i[\eta] \,,\label{barPH}
\end{equation}
where
\begin{align}
&G_1[\eta]=\int \extd x^1 G_1^{\rm bulk} \eta ,\label{G1tot}\\
&G_2[\eta]=\int \extd x^1 G_2^{\rm bulk} \eta +
\frac {\eta}{2\bar q_2} F\oB ,\label{G2tot}\\
&G_3[\eta]=\int \extd x^1 G_3^{\rm bulk} \eta +
\frac {\eta}{2\bar q_3} F\oB ,\label{G3tot}
\end{align}
and
\begin{equation}
  \label{defF}
  F\equiv \bar q_2 p_2 + \bar q_3 p_3 \,.
\end{equation}

The nice relation
\begin{equation}
H=-\sum_i G_i[\bar q_i] \label{HGi}
\end{equation}
shows that $H$ is weakly zero, because in our formulation there are boundary
terms in the Hamiltonian {\em and} in the constraints.

Among the Poisson brackets between the constraints
only non-zero ones are given explicitly:
\begin{align}
&\{ \bar P_2[\eta], G_2[\xi]\}= \frac{\eta\xi}{2 \bar q_2^2} F\oB \,,
\label{bP2G2}\\
&\{ \bar P_2[\eta], G_3[\xi]\}= -\frac{\eta\xi}{2 \bar q_2\bar q_3} F\oB \,,
\label{bP2G3}\\
&\{ \bar P_3[\eta], G_2[\xi]\}= -\frac{\eta\xi}{2 \bar q_2\bar q_3} F\oB \,,
\label{bP3G2}\\
&\{ \bar P_3[\eta], G_3[\xi]\}= \frac{\eta\xi}{2 \bar q_3^2} F\oB 
\label{bP3G3}\,,\\
&\{ G_1[\eta],G_2[\xi]\} = -G_2[\eta\xi]+\frac{\eta\xi}{\bar q_2} F\oB \,,
\label{G1G2}\\
&\{ G_1[\eta],G_3[\xi]\} = G_3[\eta\xi]-\frac{\eta\xi}{\bar q_3} F\oB \,,
\label{G1G3}\\
&\{ G_2[\eta],G_3[\xi]\} = 
-\sum_i G_i\left[ \eta\xi \frac{\partial \mathcal{V}}{\partial p_i}\right]
+\frac 12 \eta\xi F U \left( \frac{p_3}{\bar q_2} +
\frac{p_2}{\bar q_3} \right)\oB  \,.
\label{G2G3}
\end{align}
So far no specific boundary conditions have been imposed. Subsequently
several different cases are studied, all of them being consistent with
the variational principle \eqref{DbconX}, \eqref{eq:bc3}.

\subsubsection{Generic boundary}

\newcommand{\hB}{\hat{B}}

At the boundary the dilaton is given by a constant, $p_1^b$, by
assumption, while in the generic case $X^\pm\neq 0$ there. Thus we impose the boundary constraints $\hB_i$ ($E_0^\pm$ are given functions)  
\begin{align}
\hB_1[\eta]&=(p_1-p_1^b)\eta\oB\,,\label{eq:hor0.11}\\
\hB_2[\eta]&=(\bar{q}_2-E_0^-(x^0))\eta\oB\,,\label{eq:hor0.22}\\
\hB_3[\eta]&=(\bar{q}_3-E_0^+(x^0))\eta\oB\,.\label{eq:hor0.33}
\end{align}
They are obviously consistent with \eqref{DbconX}, \eqref{eq:bc3}.
Clearly $\{ \hB_i ,\hB_j\}=0$. The non-vanishing brackets with the other constraints
are
\begin{align}
&\{ \hB_2[\eta],\bar P_2[\xi]\} = \eta \xi \oB \,,\label{B2bP21}\\
&\{ \hB_3[\eta],\bar P_3[\xi]\} = \eta \xi \oB \,,\label{B3bP31}\\
&\{ \hB_1[\eta],G_2[\xi]\} = -\eta \xi p_2 \oB \,,\label{B1G21}\\
&\{ \hB_1[\eta],G_3[\xi]\} = \eta \xi p_3 \oB  \,.\label{B1G31}
\end{align}
Thus, the only constraint which obviously remains first class is
$\bar{P}_1$. The other 8 constraints $\phi_i$, $i=1..8$, yield a
matrix $M_{ij}\eta\xi\oB=\{\phi_i[\eta],\phi_j[\xi]\}$ which has
support only at the boundary. Its determinant
\begin{equation}
  \label{eq:det}
  \det{M_{ij}}=\frac{F^2}{(\bar{q}_2\bar{q}_3)^2}(p_2\bar{q}_2-p_3\bar{q}_3)^2
\end{equation}
vanishes on-shell, but has maximal rank 8 off-shell.
The gauge fixing procedure and the treatment of the delicate issue of
bulk first class constraints which become second class at the boundary
will be postponed until section \ref{se:6}. {In the previous sentence
  and the discussions below the phrase ``second class at the
  boundary'' is always understood in the sense that $M_{ij}$ has
  support only at the boundary.

\subsubsection{Horizon boundary conditions}
\label{se:hbc}

From \eqref{eq:killinghorizonws}, \eqref{eq:killinghorizon}, \eqref{DbconX} and \eqref{eq:bc3} an obvious set of boundary
constraints that corresponds to a horizon is
\begin{align}
B_1[\eta]&=(p_1-p_1^h)\eta\oB\,,\label{eq:hor0.1}\\
B_2[\eta]&=\bar{q}_2\eta\oB\,,\label{eq:hor0.2}\\
B_3[\eta]&=p_3\eta\oB\,.\label{eq:hor0.3}
\end{align}
For this choice $F\oB\approx 0$. It should be noted that on-shell $B_3$ is a consequence of $B_1$ and $B_2$, so our boundary conditions are perfectly consistent with the classical EOM \eqref{eq:EOM}.\footnote{This is no longer true if the horizon is ``stretched'' on the world sheet, i.e., if \eqref{eq:hor0.2} is replaced by \eqref{eq:hor0.22} with some ``small'' $E_0^->0$.} Again $\{ B_i ,B_j\}=0$. Non-zero brackets with the other constraints
read
\begin{align}
&\{ B_2[\eta],\bar P_2[\xi]\} = \eta \xi \oB \,,\label{B2bP2}\\
&\{ B_1[\eta],G_2[\xi]\} = -\eta \xi p_2 \oB \,,\label{B1G2}\\
&\{ B_1[\eta],G_3[\xi]\} = B_3[\eta \xi] \,,\label{B1G3}\\
&\{ B_3[\eta],G_1[\xi]\} = -B_3[\eta \xi]  \,,\label{B3G1}\\
&\{ B_3[\eta],G_2[\xi]\} = \mathcal{V}\eta \xi \oB \,.\label{B3G2}
\end{align}
Now an obstruction is encountered: the constraint $B_2$ appears in the denominator
in many places in \eqref{bP2G2}-\eqref{G2G3}. 
This reflects the well known difficulty
in constructing a canonical formulation of gravity theories
when the boundary coincides with a horizon. To be able to
proceed further one has to define a way how one treats
fractions of the constraints.
For non-extremal horizons we propose to assign the same
order to $F$, $B_2$ and to the smearing function $\xi$
corresponding to $G_2$. Then $F\xi /\bar q_2 \oB \approx 0$,
and all fractions like $F/\bar q_2$ should be considered
as finite but undetermined. These simple rules are sufficient
to make a separation between first and second
class constraints. We stress that these rules cannot be derived
from the canonical formalism, but they can be justified by
considering the behaviour of corresponding quantities near
the boundary when $x^1$ plays the role of a small parameter
($x^1=0$ is the boundary). 
Nevertheless, they are not just ad-hoc assumptions: $\bar{q}_2$ is
essentially the smearing function for $G_2$ because it appears in the
Hamiltonian as multiplier of this constraint. Thus, for consistency
one ought to require the same boundary condition for the smearing
function as for $\bar{q}_2$.
But this is not sufficient yet; we still
need some insight into the scaling behaviour of $p_3$ and $\bar{q}_2$
near a horizon in order to be able to judge whether the ratio
$p_3/\bar{q}_2$ is zero, finite or infinite. 
A straightforward analysis shows that the ratio always is finite on-shell; it may become zero for extremal horizons. Thus we have argued that for non-extremal horizons indeed  $F$, $B_2$ and the the smearing function $\xi$ corresponding to $G_2$ are of the same order.
If one accepts this rule, only one pair of second class constraints survives, namely $B_2$ and $\bar{P}_2$. All other constraints are first class.\footnote{Actually, the brackets \eqref{bP2G2}, \eqref{bP2G3} and \eqref{B2bP2} are weakly non-vanishing. However, with $\phi_i=\{G_2,G_3,B_2,\bar{P}_2\}$ the corresponding matrix $M_{ij}=\{\phi_i[\eta],\phi_j[\xi]\}$ does not have full rank, so there are only two second class constraints instead of four.}

As an alternative to 
\eqref{eq:hor0.1}-\eqref{eq:hor0.3} another boundary representing a horizon may be implemented in strict
analogy to \eqref{eq:hor0.11}-\eqref{eq:hor0.33} as
\begin{align}
\hB_1=B_1[\eta]&=(p_1-p_1^h)\eta\oB\,,\label{eq:newhor1}\\
\hB_2=B_2[\eta]&=\bar{q}_2\eta\oB\,,\label{eq:newhor2}\\
\hB_3[\eta]&=(\bar{q}_3-E_0^+(x^0))\eta\oB\,.\label{eq:newhor3}
\end{align}
It is important to realize the physical difference between the two
sets \eqref{eq:hor0.1}-\eqref{eq:hor0.3} and \eqref{eq:newhor1}-\eqref{eq:newhor3}, respectively: In the former case the
boundary is defined to be a horizon in the world-sheet
($\bar{q}_2 \approx 0$) as well as the target-space ($p_3 \approx
0$). 
In the latter case the horizon is defined in the world-sheet
only; it still may fluctuate in the target space. 
However, an off-shell
treatment of \eqref{eq:newhor1}-\eqref{eq:newhor3} turns out to be
problematic. As is obvious from \eqref{defF} $F$ does no longer
vanish at the boundary and therefore divergent results in the brackets
\eqref{bP2G2} and \eqref{bP2G3} remain unless one demands that the
smearing function of $\bar{P}_2$ vanishes at the horizon. In contrast
to $G_2$, however, such a
restriction cannot be motivated easily. Therefore we conclude that a quantum treatment of the
horizon should start from the constraints
\eqref{eq:hor0.1}-\eqref{eq:hor0.3}, which also agrees with the known technical
difficulties to define the extrinsic curvature at a horizon in the
second order formalism \eqref{eq:GDT} for the analogue of \eqref{eq:hor0.11}-\eqref{eq:hor0.33}. Henceforth exclusively \eqref{eq:hor0.1}-\eqref{eq:hor0.3} will be employed to characterise the boundary as a horizon.

\subsubsection{Bifurcation point boundary conditions}
The bifurcation point boundary con\-di\-tions $X^\pm=0$ can be obtained from the
horizon boundary conditions \eqref{eq:hor0.1}-\eqref{eq:hor0.3}, but replacing $B_2$ by
\begin{equation}
b_2[\eta]=p_2\eta\oB \,.\label{b2constr}
\end{equation}
Now only the brackets
\begin{align}
&\{ b_2[\eta],G_1[\xi]\}=\eta\xi p_2 \oB = b_2[\eta\xi]\,,\label{b2G1}\\
&\{ b_2[\eta],G_3[\xi]\}=-\eta\xi \mathcal{V} \oB \approx
-\eta\xi V \oB \,.\label{b2G3}
\end{align}
are non-vanishing. For the bifurcation point boundary conditions again $F\oB\approx 0$, and there are two pairs of second class constraints on the boundary, $b_2, G_3$ and $B_3,G_2$. 
Since $\bar q_2 \ne 0$ in general,
the canonical analysis does not require any additional assumptions.
For the special case of an extremal horizon $V\oB=0$ is valid and thus all constraints become first class. 

It should be noted that for this set of boundary conditions our assumption of a smooth boundary may not be accessible, i.e., corner contributions should be taken into account. This can be achieved most conveniently by starting from the action \eqref{act2}, but with the partially integrated version of the last boundary term. Consequently, the canonical analysis will be modified. 
In the second order formulation bifurcation point boundary conditions have been investigated thoroughly in \cite{Louko:1994tv}. 

\section{Symmetries}\label{se:5}

A natural interpretation of the canonical analysis in the previous
section is that each pair of second class constraints breaks
a gauge symmetry at the boundary, i.e., a boundary condition on
the gauge parameter is required. One can check this statement without
any use of the canonical methods by simply considering known
symmetries of the bulk action. One has to check that the action
is invariant (i.e., that the symmetry variation of the action
is zero) as well as the boundary conditions (i.e.,
that if some field is fixed on the boundary its symmetry variation
is zero). 
As the formulas are more transparent we convert within the Lagrangian formulation to the original notation of Cartan variables. 
The symmetries comprise local Lorentz transformations (with
transformation parameter $\ga$) and diffeomorphisms (with
transformation parameter $\xi^\mu$). Their infinitesimal action on the fields reads:
\begin{align}
& \de e^{\pm}_{\mu} = \pm  \ga e^{\pm}_{\mu}  +  \xi^{\nu} \partial_{\nu}
e^{\pm}_{\mu}  + \left( \partial_{\mu} \xi^{\nu}\right)e^{\pm}_{\nu}\,,
\label{variationvielbein}\\
& \de \om_{\mu} = -\partial_{\mu}\ga +  \xi^{\nu}\partial_{\nu}\om_{\mu} 
+ \left( \partial_{\mu} \xi^{\nu}\right)\om_{\nu}\,,
\label{variationconnexion}\\
& \de X =  \xi^{\nu}\partial_{\nu} X\,,
\label{variationdilaton}\\
& \de X^\pm = \pm\ga X^\pm +  \xi^{\nu}\partial_{\nu} X^\pm\,.
\label{variationtargetspace}
\end{align}
Within our choice of the boundary at fixed $x^1$ for any set of the
boundary conditions $\xi^1$ must obey
\begin{equation}
\xi^1\oB = 0\, .\label{Dirxi1}
\end{equation}
This may either be established from $\delta X\oB = 0$ or from the
condition that the gauge transformation of the action \eqref{act1} must not produce
a surface term. The latter does not yield additional
constraints: Lorentz invariance
is guaranteed by construction, 
the
remaining diffeomorphisms produce total derivatives 
along the boundary only,
\begin{align}
\delta_\xi \big(X\omega_0\oB\big) &= \partial_0\big( \xi^0 X \omega_0  \big)\,,\\
\delta_\xi \big( (\partial_0 X) \ln \frac {e_0^+}{e_0^-}
\big) &= \partial_0 \big( 
\xi^0 (\partial_0 X) \ln \frac {e_0^+}{e_0^-} \big)\,.
\end{align}
Consequently,
the action we consider is invariant under the diffeomorphism
transformations iff 
$\xi^1$ vanishes on the boundary. 
This property is shared by the EH
action with the YGH term in four dimensions.
Naturally, it is preserved by the dimensional reduction.

Now we turn to our specific choices of boundary conditions. For the generic
ones, \eqref{eq:hor0.11}-\eqref{eq:hor0.33}, inspection of the
remaining transformations immediately yields Dirichlet boundary conditions 
for all symmetry parameters. The situation
becomes more interesting in the case of a horizon. For
\eqref{eq:hor0.1}-\eqref{eq:hor0.3} it is obvious that local Lorentz
transformations and diffeomorphisms along the boundary ($\xi^0$)
are unconstrained, as they are multiplied in all relevant
transformations by quantities that have been fixed to zero. These two
results agree with the constraint analysis of the previous section; for the
generic case none of the $G_i$ remains first class at the boundary
(in the sense explained in section \ref{se:4}), for the horizon we
found that the Lorentz constraint $G_1$ and $G_3$, which may be interpreted
as diffeomorphisms along the boundary, remain strictly first class.
In the bifurcation point scenario local Lorentz
transformations are again unconstrained. However, both diffeomorphisms
must obey Dirichlet boundary conditions although $X^\pm = 0$ at the
bifurcation point, as this restriction does not hold along an
extended (one-dimensional) boundary.

It should be emphasised that the symmetry transformations can have a
non-trivial action at the boundary even if the corresponding parameter
obeys Dirichlet boundary conditions, due to the derivative terms
normal to the boundary acting on the symmetry parameters in
\eqref{variationvielbein} and \eqref{variationconnexion}. For
instance, $\partial_1\xi^1$ does not necessarily vanish at $\partial \mathcal{M}$. Similarly,
the appearance of derivatives in the constraints \eqref{eq:G1}-\eqref{eq:G3} leads to the generation of residual gauge
transformations at  $\partial \mathcal{M}$ even if they do not have
any support at the boundary or if they are second class there. 

We would like to elaborate a bit on the connection between Hamiltonian symmetries and Lagrangian symmetries for the horizon scenario since in this case we have encountered the peculiar property that the boundary constraints $B_1$, $B_3$ are first class and thus they generate gauge transformations, the meaning of which shall be clarified.
Consider first the Hamiltonian side of the picture. The gauge transformations generated by the $B_i$ read (cf.~e.g.~\cite{Henneaux:1992})
\begin{equation}
  \label{eq:hor42}
  \delta_\eps f(q_i,p_i) = \eps^j \{f(q_i,p_i),B_j\}\,,\quad j=1,3\,,
\end{equation}
where $f$ is a (differentiable) function on phase space. The only non-trivial transformations generated by the $B_i$ are
\begin{equation}
  \label{eq:hor43}
  \delta_\eps q_1 = \eps^1\,,\quad \delta_\eps q_3 = \eps^3\,.
\end{equation}
In order to understand the underlying symmetries better we consider now the Lagrangian picture, cf.~\eqref{variationvielbein}-\eqref{variationtargetspace}.
We attempt to construct the local parameters $\xi^\mu,\ga$ such that \eqref{eq:hor43} is recovered and no other quantities are being transformed. The first restriction comes from \eqref{Dirxi1}.
Consistency with $B_2$ requires
\begin{equation}
  \label{eq:hor46}
  \de e_0^- = (\partial_0 \xi^1) e_1^- \stackrel{!}{=} 0\,.
\end{equation}
Since $\partial_0$ is the derivative parallel to the boundary \eqref{eq:hor46} is fulfilled automatically without imposing any further restriction on $\xi^1$.
The constraint $B_3$ does not provide anything new neither.
Consistency with \eqref{eq:hor43} requires the vanishing of all variations besides $\de \om_1$ and $\de e_1^+$. This establishes Dirichlet boundary conditions for all symmetry variation parameters, $\ga\oB=\xi^\mu\oB=0$, as well as for their $\partial_0$ derivatives, $\partial_0\ga\oB=\partial_0\xi^\mu\oB=0$. The remaining conditions,
\begin{equation}
  \label{eq:hor48}
  \de \om_1 = -\partial_1\ga +(\partial_1\xi^1)\om_1\stackrel{!}{=}\eps^1
\end{equation}
and
\begin{equation}
  \label{eq:hor51}
  \de e_1^+ = (\partial_1\xi^0)e_0^++(\partial_1\xi^1)e_1^+\stackrel{!}{=} \eps^3\,,
\end{equation}
provide further restrictions which must be valid in any gauge (no gauge conditions have been imposed so far). Thus, they have to hold in particular in the gauge $\om_1=0=e_1^+$, which is always accessible. 
Moreover, we may assume that the horizon is located at $x^1=0$.
In such a gauge one obtains
\begin{equation}
  \label{eq:hor49}
  \ga=-\eps^1 x^1
\end{equation}
and 
\begin{equation}
  \label{eq:hor50}
  \partial_1\xi^0 = \frac{\eps^3}{e_0^+}\,.
\end{equation}
The last equation is well-defined because $e_0^+\oB\neq 0$ and the integration constant is fixed by the Dirichlet condition on $\xi^0$.

To summarise, we have established that the gauge transformations in
phase space generated by $B_1,B_3$ may be interpreted as specific
local Lorentz transformations and diffeomorphisms, respectively. Thus,
the Hamiltonian picture is consistent with the Lagrangian one, as may
have been anticipated on general grounds. This provides a further justification for our treatment of the constraint algebra in section \ref{se:hbc} (cf.~the text below \eqref{B3G2}).

\section{Reduced phase space}\label{se:6}

In order to count the number of physical degrees of freedom it is useful to take the following route: first, one may pretend that no boundaries are present, i.e., one constructs the reduced phase space for the domain $\mathcal{M}-\partial\mathcal{M}$. Then one applies a standard machinery of gauge fixing and solving constraints to construct the reduced phase space. One may then extend the results to whole $\mathcal{M}$, including $\partial\mathcal{M}$, provided the gauge fixing functions do not contradict the boundary constraints. This circumvents the challenge to deal with constraints which are first class in the bulk and second class at the boundary (in the sense explained in section \ref{se:4}): after fixing the gauge all constraints are second class and therefore may be treated on equal footing. Obviously, if one chooses a gauge which is not compatible with the boundary data inconsistencies may emerge. So one has to tread gingerly and check the consistency of all gauge fixing functions with the boundary constraints. 

For the purpose of treating all constraints on equal footing we introduce the extended Hamiltonian $H^{\rm ex}=H^{\rm tot}+\sum_i G_i[\mu_i]$, where $\mu_i$ are Lagrange multipliers for the secondary constraints $G_i$, cf.~\eqref{G1tot}-\eqref{G3tot}, and $H^{\rm tot}$ is defined in \eqref{totHam}. The canonical variables $\bar{q}_i$ now coincide with the zero components of the Cartan variables only for $\mu_i=0$. To convert the primary first class constraints $\bar P_i$ into second class constraints in the bulk we impose the gauge fixing conditions\footnote{In this context we note that the constraints $\bar{P}_i$ generate symmetries -- namely shifts of $\bar{q}_i$ -- of the extended action with the Hamiltonian $H^{\rm ex}$ but not of the original Lagrangian action \eqref{act1}. The conditions \eqref{eq:gf1}-\eqref{eq:gf3} and the constraints $\bar{P}_i$ reveal that the variables $\bar{q}_i, \bar{p}_i$ are nondynamical, as expected. These features are precisely the same as for Quantum Electrodynamics: $\bar{P}_i$, $G_i$ and $\bar{q}_i$ correspond to primary constraint, Gauss constraint and the zero component of the gauge potential, respectively.} 
\begin{align}
  \label{eq:gf1}
&  \chi_1 = \bar{q}_1 - \Om_0(x^0,x^1,q_i,p_i)\,,\\
  \label{eq:gf2}
&  \chi_2 = \bar{q}_2 - E_0^-(x^0,x^1,q_i,p_i)\,,\\
  \label{eq:gf3}
&  \chi_3 = \bar{q}_3 - E_0^+(x^0,x^1,q_i,p_i)\,.
\end{align}
Continuity requires that the limit of approaching
$\partial\mathcal{M}$ coincides with the boundary values at
$\partial\mathcal{M}$. Thus, the gauge fixing conditions
\eqref{eq:gf1}-\eqref{eq:gf3} are also valid on $\partial\mathcal{M}$.
The functions $\Om_0,E_0^-,E_0^+$ are arbitrary in principle. 
However, it is very convenient -- from a physical point of view perhaps even mandatory -- to impose $\mu_i=0$ at least at the boundary, so that boundary conditions on $\bar{q}_{2,3}$ coincide with those on $e_0^\pm$. This can be achieved easily by fixing the functions $E_0^\pm$ in \eqref{eq:gf2}, \eqref{eq:gf3} such that at $\partial\mathcal{M}$ they coincide with the corresponding boundary constraints discussed in section \ref{se:4}. We will return to this issue in more detail below, where we always will assume that the functions $\Om_0,E_0^-,E_0^+$ have been chosen such that $\mu_i\oB=0$.  

Since $\bar{q}_i, \bar{p}_i$ are nondynamical now, the reduced
phase space in the bulk may be constructed by solving $G_i=0$ together
with corresponding gauge fixings which convert them into second class constraints. A convenient set of such conditions, 
\begin{align}
  \label{eq:gf4}
&  \chi_4 = q_1 \,,\\
  \label{eq:gf5}
&  \chi_5 = q_2-1 \,,\\
  \label{eq:gf6}
&  \chi_6 = q_3 \,,
\end{align}
will be imposed on $\mathcal{M}-\partial\mathcal{M}$. 
This gauge is simple, always accessible and implies \eqref{EFmetric} for the metric.
Without loss of generality we will suppose that the boundary is placed at $x^1=0$.
At each point in the bulk one may now solve the constraints $G_i=0$ (cf.~\eqref{eq:G1}-\eqref{eq:G3}): 
\begin{align}
  \label{eq:consolv1}
 & G_2:\quad p_2=\macroT(x^0)\,,\\
  \label{eq:consolv2}
 & G_1:\quad p_1  = \macroT(x^0) x^1 + p_1^b\,,\\
  \label{eq:consolv3}
 & G_3:\quad p_3 = \frac{\macroM(x^0)-w(p_1)}{\macroT(x^0)}\exp{(-Q(p_1))}\,.
\end{align}
By continuity the solution may be extended to $\partial\mathcal{M}$. As we require the dilaton to be constant at the boundary in all prescriptions, $p_1^b$ has to be constant.
Now all $q_i,p_i,\bar{q}_i,\bar{p}_i$ are fixed and the reduced phase space has dimension zero as far as bulk degrees of freedom are concerned. There are, however, two arbitrary free functions, $\macroM(x^0)$ and $\macroT(x^0)$. Note that $\macroM(x^0)$ coincides with the Casimir function \eqref{eq:C}. Thus, in the reduced phase space there appear to remain physical degrees of freedom related to the mass $\macroM(x^0)$ and some conjugate quantity $\macroT(x^0)$. We will study now in detail different cases  to see whether these degrees of freedom are compatible with the boundary conditions.

\paragraph{Generic boundary conditions} 
The boundary constraint \eqref{eq:hor0.11} is fulfilled identically because in our solution of the constraints we have assumed that $p_1^b=\rm const.$, cf.~\eqref{eq:consolv2}. 
This property holds in all cases below. 
The functions in \eqref{eq:gf2}, \eqref{eq:gf3} have to fulfil boundary conditions consistent with the boundary constraints \eqref{eq:hor0.22}, \eqref{eq:hor0.33}. Thus, $B_2$, $B_3$ are actually redundant since they follow from continuity of the gauge fixing functions $\chi_2,\chi_3$ and consequently do not have to be counted as independent constraints. Therefore, these boundary constraints are nothing but gauge fixing constraints \eqref{eq:gf2}, \eqref{eq:gf3} evaluated at the boundary, provided the latter are chosen to be consistent with the former. If these gauge fixing functions are chosen to be inconsistent with the boundary data then a contradiction is encountered and the gauge should be discarded as inaccessible.

A formal counting
establishes $12-6-6=0$ bulk degrees of freedom, apart from possibly a finite number of global ones, which may be interpreted as being located at the boundary. Indeed, as the analysis above has shown there are two free functions $\macroM(x^0)$ and $\macroT(x^0)$ as boundary degree of freedom. The boundary phase space is two-dimensional. 

\paragraph{Horizon boundary conditions} If we choose the gauge \eqref{eq:gf1}-\eqref{eq:gf6} it turns out that in this case there remains a residual gauge freedom. The most convenient way to fix it is by replacing \eqref{eq:gf4} with
\begin{equation}
  \label{eq:yo}
  \chi_4^h = p_1 - x^1 - p_1^h\,.
\end{equation}
On a technical sidenote we remark that \eqref{eq:yo} is inaccessible in the generic case, unless the boundary constraints \eqref{eq:hor0.22}, \eqref{eq:hor0.33} are fine-tuned in a specific way. In the present case, however, the gauge is accessible because only one boundary condition on the Zweibeine, \eqref{eq:hor0.2}, is imposed.
Solving the constraints yields
\begin{align}
  \label{eq:consolv11}
 & G_1:\quad p_2=1\,,\\
  \label{eq:consolv22}
 & G_2:\quad q_1=0\,,\\
  \label{eq:consolv33}
 & G_3:\quad p_3 = (\macroM(x^0)-w(p_1))\exp{(-Q(p_1))}\,.
\end{align}
Note that on the surface of constraints the condition \eqref{eq:yo} is more restrictive than \eqref{eq:gf4} because it does not allow for a free function $\macroT(x^0)$. The geometric reason behind this simplification is the residual Lorentz symmetry discussed in section \ref{se:5}.

The function in \eqref{eq:gf2} has to fulfil Dirichlet boundary
conditions consistent with \eqref{eq:hor0.2}. For \eqref{eq:hor0.1}
the analysis of the generic case applies. However, a crucial
difference emerges due to $B_3$ as given in \eqref{eq:hor0.3}: the function $\macroM(x^0)$ is not free anymore but rather fixed by the requirement
\begin{equation}
  \label{eq:fixC}
  p_3\oB=0\quad\Rightarrow\quad \macroM(x^0)=w(p_1^h)=\rm const\,.
\end{equation}
With this condition imposed again the boundary constraints are merely a consequence of the gauge fixing functions and continuity. 
In contrast to the generic case, however, the boundary data fixes the value of the Casimir
function as well. Therefore, as anticipated by the constraint analysis
in section \ref{se:4}, the dimension of the reduced phase space is zero and
there are indeed less physical degrees of freedom if the boundary is a
horizon.

\paragraph{Bifurcation point boundary conditions}
In the gauge \eqref{eq:gf4}-\eqref{eq:gf6} again there remains some residual gauge freedom. The alternative gauge fixing \eqref{eq:yo} breaks down at the bifurcation point. Instead, by analogy to \cite{Israel:1966} one may choose 
\begin{equation}
  \chi_4^b=p_1-x^0x^1-p_1^h\,. \label{eq:bfbi1} 
\end{equation}
The bifurcation point is located at $x^0=x^1=0$, the bifurcate horizon at $x^0x^1=0$. Proceeding as above one may solve the constraints and obtains
\begin{align}
  \label{eq:consolv1bi}
 & G_1:\quad p_2=x^0\,,\\
  \label{eq:consolv2bi}
 & G_2:\quad q_1  = 0\,,\\
  \label{eq:consolv3bi}
 & G_3:\quad p_3 = \frac{\macroM(x^0)-w(p_1)}{x^0}\exp{(-Q(p_1))}\,.
\end{align}
The only crucial differences to
\eqref{eq:consolv11}-\eqref{eq:consolv33} are the possibility for $p_2$
to change sign and the appearance of a denominator $x^0$ in
\eqref{eq:consolv3bi}. The boundary constraints \eqref{eq:hor0.1} and
\eqref{b2constr} are fulfilled automatically. The remaining one,
\eqref{eq:hor0.3}, imposes the restriction \eqref{eq:fixC}. Therefore,
close to $x^0=0$ the expansion $p_3=-V(p_1^h)x^1 + \dots$
is regular, and also for the bifurcation point boundary conditions the dimension of the reduced phase space is zero.

\paragraph{Comparison with classical EOM} 
The results above may be compared with the analysis of the classical EOM in \ref{app:B} where it is also found that the boundary constraints impose certain restrictions on the solutions. It is emphasised that they arise already from the study of a single boundary.
The only arbitrariness that remains is the freedom to choose a constant $c$ in \eqref{c2c2c3c3} which corresponds to the on-shell value of the Casimir function $\macroM(x^0)$. For the boundary conditions corresponding to the horizon constraints \eqref{eq:hor0.1}-\eqref{eq:hor0.3} there are two residual gauge symmetries, 
whence the solution of the EOM is found to be unique without any free parameter to adjust. 

\paragraph{}Summarising our main results we can state that for the horizon
scenario there are fewer physical degrees of freedom in the reduced phase space as compared to generic boundary conditions, concurring with the constraint analysis in section \ref{se:4} where more gauge symmetries have been found for this case, and also in agreement with the analysis of EOM in \ref{app:B}.

\section{Conclusions and outlook}\label{se:7}

We have analysed how the presence of a horizon, interpreted as a
boundary, changes the constraint algebra and the physical phase space
compared to the presence of a general boundary. In order to reduce
clutter we have restricted the EH action to its
spherically symmetric sector. Actually, we were able to describe generic dilaton
gravity in 2D on the same footing. Furthermore, as it is advantageous to
reformulate the action in a first order form, a careful treatment of
boundaries started from the derivation of the first order form of
the YGH boundary term in section \ref{se:3}. The constraint algebra in the presence of
boundaries, studied extensively in section \ref{se:4}, led to three
cases: generic, horizon and bifurcation point boundary conditions. We
focussed on the former two, because our main interest lies in an
analysis which can be generalised to the case where matter degrees of
freedom are present -- in that case no bifurcation point is expected
to exist because the horizon does not bifurcate for a physical BH (as
opposed to an eternal BH). 

Two different sets of boundary conditions characterising a horizon
have been found. Apart from fixing the dilaton field one can impose either ``mixed'' horizon
constraints (on the world-sheet
\emph{and} on the target space) or implement the horizon solely on the world-sheet. The
latter is a limiting case of a general boundary and consequently it
can be ``stretched'' while the former cannot be deformed in this
way. Where applicable, the results for both possibilities agree.
The mixed one turned out to be the preferred set in the Hamiltonian
analysis, as for the other alternative the limiting case of a generic boundary led to
singularities we could not handle.

Somewhat surprisingly, the horizon constraint algebra revealed more first class constraints,
i.e., more generators of symmetries, as compared to the generic case. Studying the symmetries in section
\ref{se:5} showed consistency between the Lagrangian and Hamiltonian
formalism and provided also the appropriate boundary conditions for
the transformation parameters. Generically, Dirichlet boundary
conditions are required for all transformation parameters, except at
horizons or bifurcation points, where some of the symmetries (in
particular local Lorentz transformations) survive. For our preferred
set of horizon conditions
additionally diffeomorphisms along the boundary are possible.
 
Collecting the evidence obtained so far -- in particular the enhancement of symmetries at the horizon --  
we were led to study the reduced phase space in order
to decide whether or not horizon boundary conditions are special from
a physical point of view (section \ref{se:6}).  A
pivotal observation has been the consistency between the boundary
constraints and the gauge fixing functions. This allowed to regard the
former as continuation of the latter to the boundary. Due to a
convenient (Eddington-Finkelstein) choice of the gauge fixing
functions it was possible to solve all
constraints exactly. In the
generic case there remain two degrees of freedom in the physical
phase space. This is consistent with the results of Kucha{\v r} \cite{Kuchar:1994zk}, cf.~\ref{app:A}.
 However, for the horizon scenario no free function
is found
because the boundary condition defining the horizon in the target space implies the fixing
of the remaining unknown function (the Casimir function). 
As an
independent check the solutions of the EOM in presence of a boundary
are presented in \ref{app:B}. 
Full agreement with the
Hamiltonian analysis was found. 
Thus, we conclude
that {\em physical degrees of freedom are converted into gauge degrees
  of freedom on a horizon.} 

A similar statement can be found in the 2003 Erice lectures by 't
Hooft \cite{'tHooft:2004ek}. Within the ``brick wall'' model for BHs
he argued that the local gauge degrees of freedom on a horizon could represent lost information. 
This suggests to study the symmetry of the horizons and relations to the black hole
entropy with our methods. There exist various ways to count the microstates by appealing to the
Cardy formula and to recover the Bekenstein-Hawking entropy.
However, the true nature of these microstates remains unknown in this
approach, which is a challenging open problem. Many different
proposals have been made
\cite{Strominger:1996sh,Park:2001zn}, 
some of which are mutually contradicting.

From a technical point of view, the boundary Hamiltonian in the
presence of null surfaces has been constructed in the second order
formulation of 4D EH gravity in \cite{Booth:2001gx}. Also there some
differences between generic variational problems and those which
involve null surfaces were found (cf.~(26)-(27) and the surrounding
paragraphs in that work). However, no constraint analysis has been
performed and no construction of the physical phase space has been
attempted. We mention also similarities to isolated horizons
\cite{Ashtekar:2000eq}. 
In the first-order formulation of 4D EH gravity boundary conditions
have recently been studied in the Lagrangian \cite{Aros:2005jm} as well
as the Hamiltonian picture \cite{Aros:2005by}. However, sharp horizon
constraints are not implemented and a study of the constraint algebra
is not performed.

At this point we would like to compare with \cite{Carlip:2004mn} (cf.~the penultimate paragraph of section \ref{se:1}). An important difference appears to be the role of diffeomorphisms along the horizon. While in \cite{Carlip:2004mn} the vector field that generates horizon diffeomorphisms blows up, 
these transformations in our case are regular and belong to the residual gauge symmetries discussed in section \ref{se:5}. As pointed out above the horizon boundary conditions described in section \ref{se:hbc} cannot be achieved through some ``stretching procedure'', which may account for the qualitative differences between our results and \cite{Carlip:2004mn}.

Minimal \cite{Bergamin:2002ju} 
and non-minimal \cite{Bergamin:2004sr} 
supergravity extensions of the
model considered are known and it would be interesting to extend our
work to these cases. Recently it was found \cite{vanNieuwenhuizen:2005kg} that
local supersymmetry of the EH action \emph{and} of the boundary conditions
requires vanishing extrinsic curvature $K$ of the boundary. 
If one adds
the surface tension of the boundary to the action, the extrinsic
curvature
must be related to this surface tension \cite{Belyaev:2005rt}. 
It is important to find
out how these schemes work in the case of horizon boundary conditions
where the extrinsic curvature is not defined, strictly speaking.
The condition $K\oB =0$ may be of interest by itself because
in this case we do not have to fix the dilaton on the boundary to cancel
the corresponding part of (\ref{bvari}). 

From our experience with the constraints in the bulk alone
\cite{Grumiller:2002nm}, the
analysis will not differ very much after matter has been
included. There are, however, some relevant changes regarding the
boundary constraints. A crucial observation in this context is that
the horizon condition in the target space
\eqref{eq:killinghorizon} still describes a trapping horizon! One can
derive this statement by analysing the EOM \eqref{eq:EOM} suitably
extended to include matter (for spherically reduced gravity it is
particularly transparent because the fields $X^\pm$ correspond to the
expansion spin coefficients). Thus, the corresponding horizon constraint \eqref{eq:hor0.3}
still may be imposed. Moreover, the coordinate $x^1$ can again be
chosen constant along the boundary.  However, the trapping horizon
will not be a Killing horizon in general and thus the---in the
matterless case equivalent---world sheet condition
\eqref{eq:killinghorizonws} is no longer true. Accordingly, the
(Killing) horizon condition
\eqref{eq:hor0.2} has to be replaced by \eqref{eq:hor0.22}. Moreover,
the dilaton field need not be constant along the boundary anymore,
which enforces a further generalisation. Note that the boundary no longer is a null-surface and
therefore the presence of the logarithmic contribution to the YGH term in \eqref{act1} no
longer is dangerous. 
In the present derivation the horizon condition on the target space
\eqref{eq:hor0.3} plays a central role in the symmetry enhancement on
the horizon. As it still holds after matter has been included it may be
expected that the physical phase space is again smaller as compared to
the case of generic boundary conditions. It would be nice to verify this
conjecture.

Any degrees of freedom which are observed in the reduced phase space
formalism should be seen also in the path integral. Actually, we already have made some progress towards generalising the path integral for 2D gravity
\cite{Kummer:1997hy,Grumiller:2000ah,Grumiller:2002nm} 
in the presence of  
boundaries. What is missing is the BRST charge, which is a technically demanding calculation, but no essential difficulties are expected.
With a path integral at hand one may study, for example, the interaction
between virtual black holes \cite{Grumiller:2000ah} 
and boundaries.

\ack

We thank H.~Balasin for many helpful discussions.
DG is grateful to S.~Carlip for explanations of his work during a
conference in Sardinia and to T.~Strobl for discussions on \cite{Gegenberg:1997de,Strobl:1999wv}.

LB has been supported by project P-16030-N08 of the Austrian Science Foundation (FWF). DG has been supported by an Erwin-Schr\"odinger fellowship, project J-2330-N08 of the  Austrian Science Foundation (FWF). DVV has been supported by project BO 1112/12-2 of the German Research Foundation (DFG).

DG and DVV acknowledge the hospitality at the Vienna
University of Technology 
during the preparation of this work. DG also acknowledges the
hospitality at the University of Jena.
LB acknowledges the hospitality at the University of Leipzig
during the preparation of this work and would like to thank R.~and
M.~Poncet for a wonderful stay in Termes (Ard\`eche, FR) during the
final stage of this work.
Travel support of DG is due to project P-16030-N08 of the Austrian Science Foundation (FWF).
Travel support of DVV is due to the Multilateral Research Project ``Quantum gravity,
cosmology and categorification'' of the Austrian Academy of Sciences and the
National Academy of Sciences of the Ukraine.

\begin{appendix}

\section{Analysis of EOM}\label{app:B}

Variation of \eqref{eq:FOG} in the bulk yields the EOM\\
\parbox{0.9\linewidth}{\begin{align*}
 & \extd X+X^{-}e^{+}-X^{+}e^{-}=0\, , \\
 & (\extd\pm \omega )X^{\pm }\mp \mathcal{V}e^{\pm }=0\, , \\
 & \extd\omega +\epsilon \frac{\partial \mathcal{V}}{\partial X}=0\, , \\
 & (\extd\pm \omega )e^{\pm }+\epsilon \, \frac{\partial \mathcal{V}}{\partial X^{\mp }}=0\, ,
\end{align*}}\hfill\parbox{8mm}{\begin{align}\label{eq:EOM}\end{align}}\\
in covariant form.
To simplify their analysis let us impose the gauge conditions \eqref{eq:gf4}-\eqref{eq:gf6}, i.e., $\om_1=0=e_1^+$ and $e_1^-=1$.
These gauge conditions leave unbroken diffeomorphisms and local Lorentz transformations provided the transformation parameters fulfil
\begin{equation}
  \label{eq:unbroken}
  \xi^0=\xi^0(x^0)\,,\qquad \ga + \partial_1\xi^1 = 0\,,\qquad \partial_1\ga=0\,.
\end{equation}

From the Hamiltonian point of view there are three sets of EOM. The first one is
obtained by varying the bulk part of the action (\ref{act1}) 
with respect to $\bar q_i$.
\begin{align}
&\partial_1 p_1 -p_2 =0,\label{eom11}\\
&\partial_1 p_2 = 0,\label{eom12}\\
&\partial_1 p_3 + \mathcal{V} (p_2p_3,p_1)=0.\label{eom13}
\end{align}
These equations coincide with the bulk parts of corresponding 
constraints $G_i$ after the gauge conditions \eqref{eq:gf4}-\eqref{eq:gf6} have been taken into account. The second set is
produced by the variations with respect to $q_i$,
\begin{align}
&\partial_0 p_1 - p_2 \bar q_2 + p_3 \bar q_3 =0,\label{eom21}\\
&\partial_0 p_2 + p_2 \bar q_1 - \bar q_3 \mathcal{V}=0,\label{eom22}\\
&\partial_0 p_3  -p_3 \bar q_1 + \bar q_2 \mathcal{V} =0,\label{eom23}
\end{align}
and the last set is generated by $p_i$
\begin{align}
&\partial_1 \bar q_1 -\bar q_3 \frac{\partial\mathcal{V}}{\partial p_1} =0,
\label{eom31}\\
&\partial_1 \bar q_2 + \bar q_1 - \bar q_3 U p_3=0,\label{eom32}\\
&\partial_1 \bar q_3 - \bar q_3 U p_2 =0. \label{eom33}
\end{align}

\paragraph{Generic boundary}
The boundary conditions are given by \eqref{eq:hor0.11}-\eqref{eq:hor0.33}.
There is no residual gauge freedom associated with these boundary
conditions because $\xi^1$ fulfils Dirichlet boundary conditions and neither $\xi^0$ nor $\ga$ (nor any combination thereof) may be chosen freely without violating the boundary conditions.
The solution of \eqref{eom11}-\eqref{eom13} is provided in \eqref{eq:consolv1}-\eqref{eq:consolv3}. Note the emergence of two integration functions $\macroM(x^0)$ and $\macroT(x^0)$.
We solve some of the remaining EOM in the order \eqref{eom33}, \eqref{eom22}, \eqref{eom21}:
\begin{align}
&\bar q_3 = E_0^+ (x^0) e^Q \,, \label{gsolbq3}\\
&\bar q_1 = \frac {\bar q_3 \mathcal{V}}{\macroT} -
\frac{\dot{\macroT}}{\macroT} \,,\label{gsolbq1}\\
&\bar q_2 = \macroT^{-1} ( p_3 \bar q_3 + x^1 \dot{\macroT}) \,,\label{gsolbq2}
\end{align}
where $\dot{\macroT}=\partial_0 \macroT$. 
The lower integration limits in the functions
\begin{align}
&Q(p_1)\equiv \int_{p_1^h}^{p_1} \extd y\, U(y),\label{defQ}\\
&w(p_1)\equiv \int_{p_1^h}^{p_1} \extd y\, V(y)e^{Q(y)}.\label{defw}
\end{align}
have been chosen for convenience. 
All integration constants are captured by the two free functions $\macroT$ and $\macroM$ and the boundary value $E_0^+$.
Thus, at this stage all canonical variables are determined uniquely up to $\macroT$, $\macroM$ and $E_0^+$.

The equations (\ref{eom31}) and (\ref{eom32}) are satisfied automatically,
but equation (\ref{eom23}) yields the condition
\begin{equation}
\dot{\macroM}=0 \quad\Rightarrow\quad M(x^0) =c\,,\label{c2c2c3c3}
\end{equation}
with some constant $c$. The boundary condition (\ref{eq:hor0.22})
fixes the remaining integration function:
\begin{equation}
 \macroT (x^0) = \sqrt{ c\, \frac{E_0^+(x^0)}{E_0^-(x^0)}} \,.
\label{c2x0}
\end{equation}
The only arbitrariness that remains is the freedom to choose the
constant $c$. This quantity corresponds to the on-shell value of the
Casimir function \eqref{eq:C}. Its sign is fixed by
the requirement of reality of $\macroT$, in particular if $E_0^+$ and
$E_0^-$ are (semi-)positive as assumed in the main text then $c$ is
(semi-)positive as well.

\paragraph{Horizon boundary conditions}
We impose the boundary conditions corresponding to the constraints
(\ref{eq:hor0.1})-(\ref{eq:hor0.3}). There are two residual gauge
symmetries corresponding to $\xi^0(x^0)$ (diffeomorphisms) and
$\ga (x^0)$ (Lorentz). 
The general solution of (\ref{eom12}) reads $p_2=\macroT(x^0)$. By an 
$x^0$-dependent finite Lorentz transformation one can always achieve
$\macroT(x^0)=1$, so that
\begin{equation}
p_2=1 .\label{hsolp2}
\end{equation}
Then the equations (\ref{eom11}) and (\ref{eom13}) have unique
solutions satisfying the boundary conditions defined by 
(\ref{eq:hor0.1}) and (\ref{eq:hor0.3}):
\begin{align}
&p_1=x^1 + p_1^h,\label{hsolp1}\\
&p_3=-e^{-Q} w,\label{hsolp3}
\end{align}
where $x^1=0$ corresponds to the boundary and the definitions
\eqref{defQ}, \eqref{defw} have been used. It is pivotal that no ambiguity is present in
\eqref{hsolp3} because $p_3$ obeys Dirichlet boundary conditions.

Next we use the residual gauge freedom associated with $\xi^0(x^0)$ to
make
\begin{equation}
(\bar q_3 = e_0^+)\oB =1 .\label{eopoB}
\end{equation}
Then (\ref{eom33}) is solved uniquely,
\begin{equation}
\bar q_3 = e^{Q} .\label{hsolbq3}
\end{equation}
The quantities $\bar q_1$ and $\bar q_2$ can be found from (\ref{eom22}) and (\ref{eom21}),
respectively:
\begin{align}
&\bar q_1=e^Q\mathcal{V} ,\label{hsolbq1}\\
&\bar q_2=e^Q p_3.\label{hsolbq2}
\end{align}
The remaining equations (\ref{eom23})-(\ref{eom32})
and the boundary condition (\ref{eq:hor0.2}) are satisfied automatically.
The solution is fixed uniquely by the boundary conditions. 

\paragraph{Bifurcation point boundary conditions}
Solving the EOM in the same manner as before and adopting the choices
\eqref{defQ} and \eqref{defw} yields
\begin{align}
  & p_1=x^0x^1+p_1^h\,,\\
  & p_2=x^0\,,\\
  & p_3=- \frac{w}{x^0}e^{-Q}\,,\\
  & \bar{q}_1 = -\frac{E_0^+(x^0)\left(w U -w^\prime\right)+1}{x^0}\,,\\
  & \bar{q}_2 = \frac{x^0x^1-E_0^+(x^0)w}{(x^0)^2}\,,\\
  & \bar{q}_3 = E_0^+(x^0)e^{Q}\,.
\end{align}
 The residual local Lorentz transformations can be exploited to make $E_0^+(x^0)=E_0^+=\rm const.$ Regularity of the coordinate system at $x^0=x^1=0$ requires
 \begin{equation}
   \label{eq:regularity}
   E_0^+=\frac{1}{w^\prime(p_1^h)}\,.
 \end{equation}
The quantity $w^\prime(p_1^h)$ is essentially surface gravity, so for non-extremal horizons $E_0^+$ is well-defined. Again the solution is fixed uniquely by the boundary conditions.

\paragraph{Alternative horizon boundary conditions}

In contrast to the constraint algebra, Lagrangian symmetries and
the EOM can be analysed for the
alternative horizon prescription \eqref{eq:newhor1}-\eqref{eq:newhor3}
as well. At first glance the situation appears to be different, as
there remains only one residual gauge transformation $\xi^0(x^0)$ provided the other transformation parameters fulfil
\begin{equation}
  \label{eq:newhor4}
  \ga E_0^+ (x^0) + \partial_0 \left(\xi^0 E_0^+ (x^0) \right) =
  0\,,\quad \xi^1=x^1\ga\, .
\end{equation}
However, this is sufficient to achieve \eqref{hsolp2} and thus the analysis above applies also to this case, with the sole replacement of 1 by $E_0^+$ in the right hand side of \eqref{eopoB}. The reason for this equivalence is that on-shell the constraint $p_3=0$ is implied by the boundary constraint $\bar{q}_2=0$.

\section{Abelianised constraints: Relation to Kucha\v{r}}\label{app:A}

In \cite{Kuchar:1994zk} Kucha\v{r} 
constructed a reduced phase space for the Schwarzschild BH which
comprises the BH mass and extrinsic time as canonical
variables, 
in a way very different from the main text. Here we make 
contact with Kucha\v{r}'s work. To this end it is very convenient to further simplify the constraint algebra \eqref{eq:consalg}. 

In the absence of boundaries within the PSM formulation it has been realized that the constraint algebra may be abelianised not only locally but in a certain patch which covers e.g.~the whole exterior of a BH \cite{Schaller:1994np,Strobl:1999wv}.
Following appendix E of \cite{Grumiller:2001ea} we define a new set of constraints
\begin{align}
  \label{eq:e1}
  G_1^c &:= G_1 \,,\\
  \label{eq:e2}
  G_2^c &:= e^{-Q(p_1)}\frac{p_3G_2-p_2G_3}{2p_2p_3}\,,\\
  \label{eq:e3}
  G_3^c &:= e^{Q(p_1)}\left(\mathcal{V}G_1+p_3G_2+p_2G_3 \right)=\partial_1 \mathcal{C}\,,
\end{align}
where the $G_i$ are the bulk constraints defined in \eqref{eq:G1}-\eqref{eq:G3} and $\mathcal{C}$ is defined in \eqref{eq:C}. This redefinition of constraints is well-defined in a patch where $p_2p_3\neq 0$, which by virtue of \eqref{eq:killinghorizon} requires the absence of a Killing horizon in the whole patch. Thus, the abelianised constraints \eqref{eq:e1}-\eqref{eq:e3} are not useful for horizon boundary conditions, but they may be employed for generic boundary conditions in the outside region of a BH.
It is convenient to introduce a new set of canonical variables:
\begin{align}
  \label{eq:e5}
  q_1^c &:= q_1-\frac{q_2p_2+q_3p_3}{2p_2p_3}\,\mathcal{V}\,, & p_1^c := p_1\,,\\
  \label{eq:e4}
  q_2^c &:=q_2p_2-q_3p_3\,, & p_2^c :=\frac12\ln{\frac{p_2}{p_3}}\,,\\
  \label{eq:e6}
  q_3^c &:= e^{-Q(p_1)}\frac{q_2p_2+q_3p_3}{2p_2p_3}\,, & p_3^c := p_2p_3e^{Q(p_1)}+w(p_1)=\mathcal{C}\,.
\end{align}
Both the Jacobian in the transformation of the constraints and the Jacobian in the canonical coordinate transformation equal to unity. One may easily check that
\begin{equation}
  \label{eq:e9}
  \{q_i^c,{p_j^c}^\prime\}=\de_{ij}\de\,,\quad \{q_i^c,q_j^c\}=0=\{p_i^c,p_j^c\}\,.
\end{equation}
The constraints $G_i^c$ in terms of these variables are very simple:
\begin{align}
\label{eq:e10} & G_1^c=\partial_1p_1^c-q_2^c\,,\\
\label{eq:e11} & G_2^c=\left(\partial_1p_2^c+q_1^c\right)e^{-Q(p_1^c)}\,,\\
\label{eq:con} & G_3^c=\partial_1p_3^c\,.
\end{align}
Their interpretation is as follows: $G_1^c$ generates Lorentz transformations (being canonically conjugate to the Lorentz angle $p_2^c$), $G_2^c$ generates radial translations (essentially being canonically conjugate to the dilaton $p_1^c$), $G_3^c$ generates time translations (being the derivative of the mass $p_3^c$ which, roughly speaking, is canonically conjugate to ``time''). The constraints are abelian up to boundary terms,\footnote{This seems to be the proper place to make a parenthetical remark regarding the stability of the constraint algebra against deformations due to addition of matter: the abelian algebra \eqref{eq:e12} is very unstable in the sense that addition of minimally coupled matter already deforms it in a complicated way. Thus, for considerations of quantum effects or inclusion of matter it seems to be of little use. The ``original'' constraint algebra with the $G_i$ is stable against addition of minimally coupled matter and even for non-minimally coupled matter only the structure function $C_{23}{}^1$ is changed.}
\begin{equation}
  \label{eq:e12}
  \{G_i^c,{G_j^c}^\prime\}=0\,,\quad\forall\,\,i,j\,.
\end{equation}
It is possible to make them abelian even in the presence of a boundary by adding to the smeared constraint $G_2^c[\xi]$ a boundary term $\xi p_2^c \exp{(-Q(p_1^c))}\oB$.

The Hamiltonian action now reads
\begin{equation}
  \label{eq:e13}
  S^{(c)}=\int_\mathcal{M}\extd^2x \left(p_i^c\dot{q}_i^c+\bar{q}_i^cG_i^c\right) + S_B\,,
\end{equation}
where $S_B$ denotes a boundary term. The new Lagrange multipliers may be expressed in terms of the old ones by requiring equivalence of the bulk Hamiltonians, $\bar{q}_i G_i=\bar{q}_i^cG_i^c$.

With the experience of the main text it is convenient to make a
short-cut rather than repeating the more elaborate analysis of
sections \ref{se:3}-\ref{se:6}: because the last term in
\eqref{act1} vanishes for a constant dilaton we know already that
arguments of functional differentiability (which may be invoked to
establish the first boundary term in \eqref{act1}) are sufficient if the dilaton is constant at the boundary, which again will be assumed to be a surface of $x^1=\rm const.$ Thus, it suffices to declare which quantities are held fixed at the boundary in order to construct $S_B$. The first condition is constancy of the dilaton,
\begin{equation}
  \label{eq:appA1}
  \de p_1^c\oB=0\,.
\end{equation}
For the second condition we note that the Lagrange multiplier
\begin{equation}
  \label{eq:appA2}
  \bar{q}_2^c=e^{Q(p_1)}(\bar{q}_2p_2-\bar{q}_3p_3)\,,
\end{equation}
vanishes on-shell. Thus, a choice consistent with the classical EOM is
\begin{equation}
  \label{eq:appA3}
  \bar{q}_2^c\oB=0\,.
\end{equation}
No contribution to $S_B$ has been produced by the conditions \eqref{eq:appA1} and \eqref{eq:appA3}. If one would like to have $S_B=0$ then the relation
\begin{equation}
  \label{eq:e14}
  \bar{q}_3^c\de p_3^c = 0\,,
\end{equation}
has to be fulfilled. The quantity
\begin{equation}
  \label{eq:appA5}
  \bar{q}_3^c=e^{-Q(p_1)}\frac{\bar{q}_2p_2+\bar{q}_3p_3}{2p_2p_3}
\end{equation}
may be interpreted as lapse function and there is no geometric reason why it should vanish. Fixing $p_3^c$ at the boundary means that the mass must not fluctuate there. 
This is a valid choice, but it is something that Kucha\v{r} wants to avoid in his approach. Therefore, we consider now another possibility.

We may choose in \eqref{eq:e13} 
\begin{equation}
  \label{eq:e15}
  S_B=-\int_{\partial\mathcal{M}}\extd x^0 \bar{q}_3^c p_3^c\,.
\end{equation}
This coincides with Kucha\v{r}'s result if we call $p_3^c=M$ and
$\bar{q}_3^c=N$. With this addition the variation at the boundary now
yields instead of \eqref{eq:e14}
\begin{equation}
  \label{eq:e16}
   p_3^c \de\bar{q}_3^c = 0\,.
\end{equation}
This means that in order to get a nonvanishing mass ($p_3^c\neq 0$) we were forced to assume that the variation of the lapse vanishes at the boundary, $\de\bar{q}_3^c=0$.
To avoid this requirement one can now proceed as Kucha\v{r} does, i.e., to introduce a ``proper time function'' $\tau$ instead of $\bar{q}_3^c$ given by 
\begin{equation}
  \label{eq:trick}
  \dot{\tau}=-\bar{q}_3^c\,.
\end{equation}
 Consequently, one may keep variations of $\tau$ and $\bar{q}_3^c$ arbitrary at the boundary and obtains the boundary action 
 \begin{equation}
   \label{eq:bac}
   S_B=\int_{\partial\mathcal{M}}\extd x^0 \dot{\tau} M\,.
 \end{equation}
After going to the reduced phase space these are the only remaining physical degrees of freedom, and therefore $\dot{p}_3^c=\dot{M}=0$, i.e., mass conservation is implied on-shell (see section 3.5 in \cite{Kuchar:1994zk}). In principle one could apply the ``trick'' of introducing an additional derivative as in \eqref{eq:trick} several times, thereby producing an arbitrary number of ``time'' derivatives in the boundary action. However, after the second time mass conservation no longer is implied on-shell, so in this sense the action \eqref{eq:bac} is preferred. 


\end{appendix}

\section*{References}

\providecommand{\href}[2]{#2}\begingroup\raggedright\endgroup





\end{document}

%% file: macros.tex
\newcommand{\beqs}{\begin{equation*}}
\newcommand{\beq}{\begin{equation}}

\newcommand{\eeqs}{\end{equation*}}
\newcommand{\eeq}{\end{equation}}

\newcommand{\beqas}{\begin{eqnarray*}}
\newcommand{\beqa}{\begin{eqnarray}}

\newcommand{\eeqas}{\end{eqnarray*}}
\newcommand{\eeqa}{\end{eqnarray}}

\newcommand{\eq}[2]{\begin{equation} #1 \label{#2} \end{equation}}

\newcommand{\eps}{\varepsilon}

\newcommand{\ga}{\gamma}
\newcommand{\de}{\delta}
\newcommand{\om}{\omega}

\newcommand{\la}{\lambda}
\newcommand{\si}{\sigma}

\newcommand{\Om}{\Omega}

\newcommand{\blist}{\begin{itemize}}

\newcommand{\elist}{\end{itemize}}

\providecommand{\href}[2]{#2}

\DeclareFontFamily{OT1}{rsfs}{}
\DeclareFontShape{OT1}{rsfs}{m}{n}{ <-7> rsfs5 <7-10> rsfs7 <10->rsfs10}{} 
\DeclareMathAlphabet{\mycal}{OT1}{rsfs}{m}{n}